\documentclass{article}
\usepackage{spconf,amsmath,graphicx}
\usepackage{booktabs}
\usepackage{multirow}
\usepackage{verbatim}
\usepackage{url}
\usepackage{subfigure}
\usepackage{graphicx}
\usepackage{tabularx}
\usepackage{bm}
\usepackage{enumitem}
\usepackage{algorithm, algorithmic}
\usepackage{amssymb}
\title{Using Multiple Reference Audios and Style Embedding Constraints for Speech Synthesis}
\name{Cheng Gong$^1$, Longbiao Wang$^{1*}$, Zhenhua Ling$^{2}$, Ju Zhang$^3$, Jianwu Dang$^{1,4}$\thanks{* Corresponding Author.}}
\address{$^1$Tianjin Key Laboratory of Cognitive Computing and Application,\\ College of Intelligence and Computing, Tianjin University, Tianjin, P.R.China\\
$^2$National Engineering Laboratory for Speech and Language Information Processing,\\
University of Science and Technology of China, Hefei, P.R.China\\
$^3$Huiyan Technology (Tianjin) Co., Ltd\\
$^4$Japan Advanced Institute of Science and Technology, Ishikawa, Japan\\
\small \{gongchengcheng, longbiao\_wang\}@tju.edu.cn, zhling@ustc.edu.cn}
\begin{document}
\ninept
\maketitle
\begin{abstract}
The end-to-end speech synthesis model can directly take an utterance as reference audio, and generate speech from the text with prosody and speaker characteristics similar to the reference audio.
However, an appropriate acoustic embedding must be manually selected during inference. Due to the fact that only the matched text and speech are used in the training process, using unmatched text and speech for inference would cause the model to synthesize speech with low content quality. In this study, we propose to mitigate these two problems by using multiple reference audios and style embedding constraints rather than using only the target audio. Multiple reference audios are automatically selected using the sentence similarity determined by Bidirectional Encoder Representations from Transformers (BERT). 
In addition, we use ``target'' style embedding from a pre-trained encoder as a constraint by considering the mutual information between the predicted and ``target" style embedding.
The experimental results show that the proposed model can improve the speech naturalness and content quality with multiple reference audios and can also outperform the baseline model in ABX preference tests of style similarity.

\end{abstract}
\section{Introduction}
\label{sec:intro}
The goal of text-to-speech (TTS) is to synthesize intelligible and natural speech.
Recent advances TTS have significantly improved the naturalness of synthetic speech \cite{wang2017tacotron,shen2018natural,li2019neural}.
Naturalness largely depends on the expressiveness of the synthesized voice, which is determined by multiple characteristics, such as content, timbre, prosody, emotion, and style \cite{tan2021survey}. 
Despite having important applications, such as conversational assistants and long-form reading, the development of expressive TTS is still considered an important open problem.

To deliver true human-like speech, a TTS system must learn to model prosody. Researchers have addressed this problem by providing an additional reference speech signal to control the style of the generated speech \cite{skerry2018towards,wang2018style,akuzawa2018expressive,zhang2019learning,ma2019generative}. But these methods still have two problems when synthesizing a sample: 1) one has to manually select an appropriate acoustic embedding, which can be challenging, and 2) during the training phase, the reference utterance and the input text are paired (i.e., the text is the transcription of the reference utterance). However, the input reference utterance and text are not paired during testing. Because the unpaired inputs are never seen during training, the generated voice has low speech naturalness and speech content quality. 

One solution for prosody selection could be to remove the manually selection process by predicting the style from the text.
In \cite{stanton2018predicting}, the Text-Predicted Global Style Token (TP-GST) architecture learned to predict stylistic renderings from text alone, requiring neither explicit labels during training nor auxiliary inputs for inference. 
Similarly, Karlapati et al. \cite{karlapati2021prosodic} proposed a method to sample from this learned prosodic distribution using the contextual information available in the text.
However, it is difficult to predict the latent style from text alone because the process of style embedding is an entangled representation of prosody and unknown underlying acoustic factors. 
Instead of using only text, Tyagi et al. \cite{tyagi2020dynamic} proposed an approach that leverages linguistic information to drive the style embedding selection of such systems. In this method, a sentence is selected as the reference audio if its linguistic similarity is similar to that of the target. However, only one audio, rather than multiple audios, was selected for style embedding.

Moreover, if the network architecture is not carefully designed, the generated voice is influenced by the content of the reference utterance.  
To tackle this content leakage problem, previous studies have proposed several methods that employ unpaired data training.
For example, Liu et al. \cite{liu2018improving} proposed to mitigate the problem by using the unmatched text and speech during training, utilizing the automatic speech recognition (ASR) accuracy of an end-to-end ASR model to guide the training procedure.
In addition, using the unpaired data, Ma et al. \cite{ma2018neural} introduced an end-to-end TTS model by combining a pairwise training procedure, an adversarial game, and a collaborative game into one training scheme. 

In addition to methods that use unpaired data, a number of recently proposed methods improve the content quality of synthesized speech by adding constraints to the style embeddings.
Ideally, the style embedding should not be able to reconstruct the content vector (i.e., there should be no information about the content in the style embedding). To this end, Hu et al. \cite{hu2020unsupervised} minimized the mutual information (MI) between the style and content vectors. In addition, MI has been applied to other speech synthesis tasks, such as learning disentangled representations of the speakers and the languages they speak \cite{xin2021disentangled}, and maximizing the mutual information between the text and acoustic features to strengthen the dependency between them \cite{liu2019maximizing}. In this study, we also design a constraint for style embedding using mutual information.

Combining the aforementioned ideas of using text information to predict style embeddings and using unpaired data to improve speech content quality,
we propose multi-reference TTS (MRTTS). This model is trained using multiple automatically selected and weighted reference audio to generate expressive speech.
The contributions of this paper are as follows. \emph{I)} We present an approach that uses automatically selected and weighted multi-reference audios for speech style modeling. \emph{II)} For the proposed method, we design a constraint for learned style embeddings using mutual information. \emph{III)} 
To the best of our knowledge, this work is the first attempt to improve expressive speech quality using multiple reference audios rather than only one target audio.
The objective and subjective evaluation results demonstrate that MRTTS exhibits superior performance compared to the baseline model in terms of speech naturalness, speech content quality, and style similarity.

%The remainder of this paper is structured as follows. In Section 2, we describe our proposed method. Section 3 presents the experimental conditions and the results of the subjective and objective experiments. Section 4 concludes the paper with our findings and our future work.

%The remainder of this paper is structured as follows. In Section 2, we describe our proposed method. Section 3 presents the experimental conditions and the results of the subjective and objective experiments. Section 4 concludes the paper with our findings and our future work.
%two wav to solve prosody select, predict, bert. 训练

%content leakage, unpaired data. disentangled
\section{PROPOSED MRTTS MODEL}
\begin{figure}[t]
\centering
\subfigure[Linguistics-driven multiple reference selection.]{
\includegraphics[width=0.48\textwidth]{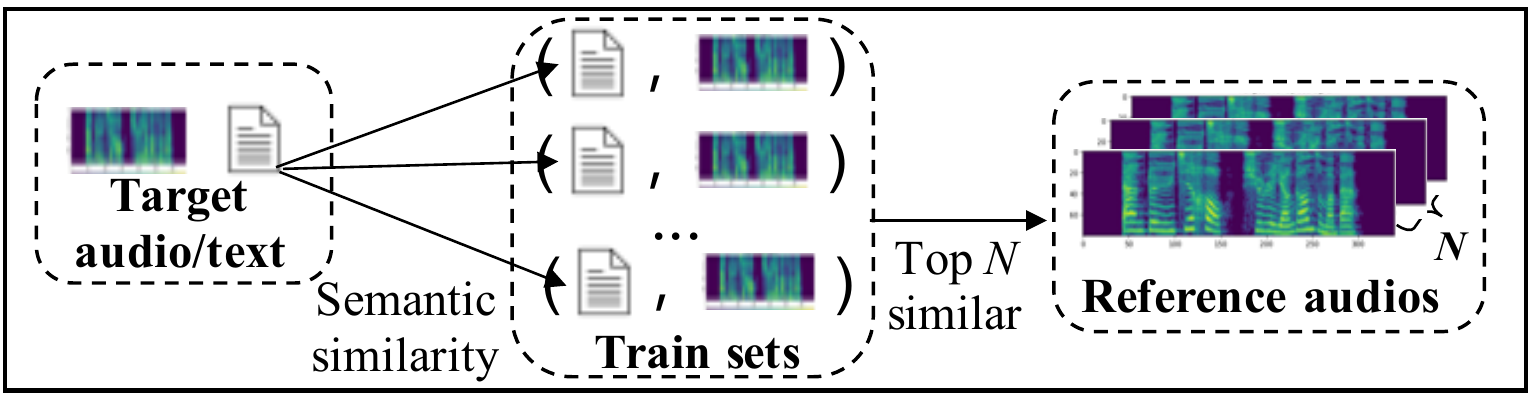}
%\caption{fig1}
}
\quad
\subfigure[Model diagram. The blue dashed box means style constraint module which uses a pre-train style encoder to constrain the style embedding during the training process.]{
\includegraphics[width=0.48\textwidth]{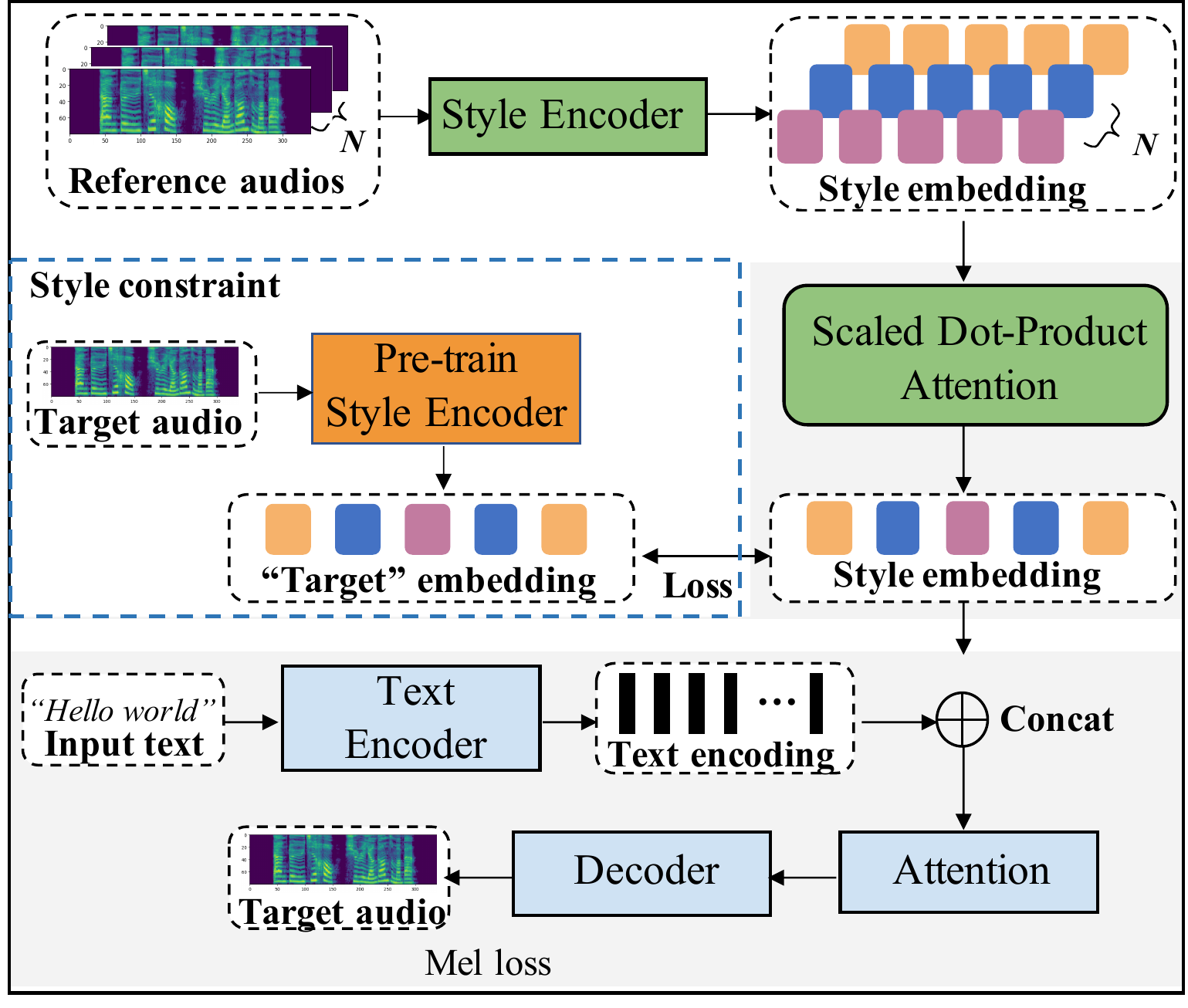}
}
\vspace{-0.5cm}
\caption{Proposed MRTTS model.}
\label{fig:model structure}
\vspace{-0.4cm}
\end{figure}

%MINE？
The proposed method, shown in Fig. \ref{fig:model structure}, is based on the end-to-end TTS architecture Tacotron2. For the text encoder input $x$, we used the character sequence of the normalized text for training. 
All the style encoders learned to model prosody from one or multiple reference audio with the same settings as the Global Style Token (GST) module.
Multiple $N$ reference audios were selected based on the corresponding text similarity as determined by BERT.
For multiple style embeddings obtained from multiple reference audios, we used the scaled attention method to obtain the final style embedding.
%Training methods can be divided into supervised and unsupervised methods, and supervised methods require an additional pre-trained model to extract the "target" style embedding.
\subsection{Linguistics-driven multiple references selection}
%{Automatically select and weight multiple reference audios}
The objective of this selection step is to select reference audios from training sets for training and inference. Unlike GST, which uses the target audio as the reference audio, we selected the reference audio to be the audio that has a semantic similar to the target, as in \cite{tyagi2020dynamic}. We defined the semantic distances as the cosine similarity between the two sentence-level embeddings, and we selected the reference audio to be the audio for which the semantic distance is closest to the target audio, as shown in Fig. \ref{fig:model structure}(a).
%公式表示top

To generate sentence embedding, we used BERT \cite{devlin2019bert,xiao2018bertservice} because it is one of the best pre-trained models and it produces state-of-the-art results on a large number of NLP tasks. We used word representations from the uncased base (12-layer) model without fine tuning. Sentence-level representations were achieved by averaging the second to last hidden layers for each token in the sentence. We did not use ``[CLS]" because it acts as an ``aggregate representation" for classification tasks and it is not the best choice for quality sentence embeddings.

\subsection{Multi-references based acoustic modeling}
%\textbf{Attention module.}
From the selected $N$ reference audios, the style encoder obtains $N$ style embedding $S
=[s_1,s_2,...,s_n]$. The attention query $Q$, key $K$, value $V$, and final style embedding $E$ are calculated as
%\begin{center}
\begin{eqnarray}
\setlength{\abovedisplayskip}{0.5pt}
\setlength{\belowdisplayskip}{0.5pt}
\begin{aligned}
  \bm{Q} = \bm{Q^{'}W_q}, \quad  \bm{K} = \bm{SW_k}, \quad  \bm{V} = \bm{SW_v},\\
  \bm{E} = \text{A}(\bm{Q},\bm{K}, \bm{V})=\text{softmax}(\frac{f(\bm{Q})f(\bm{K})^T}{\sqrt{d_m}})f(\bm{V}),
  %T_t &=& C_t * N_f/M_f      \\
  %T_l &=& C_l * F_l * N_f  \nonumber
  \end{aligned}
  \label{eq1}
\end{eqnarray}
%\end{center}
respectively, where $\bm{Q^{'}}$ is an embedding in which the values are randomly initialized and automatically learned by backpropagation, and all $W$ are linear projection matrices.

Here, attention is not used to learn the alignment. Instead, it learns a similarity measure between the $\bm{Q^{'}}$ vector and each style embedding $s_i$ in $S$. 
The weighted sum of the style embedding $S$, which we call the final style embedding $E$, is passed onto the text encoder outputs for conditioning in every time step.

Without the style embedding constraints, as shown in Fig. \ref{fig:model structure}(b), the style encoder is jointly trained with the rest of the model, driven only by the Mel reconstruction loss $\mathcal{L}_{mel}$ from the Tacotron2 decoder, as in \cite{wang2018style}. Thus, the style encoder does not require any explicit style or prosody labels.

\subsection{Model training with style embedding constraints}
%Without style constraint, as shown in Fig. \ref{fig:model structure} (b), the style encoder is jointly trained with the rest of the model, driven only by the Mel reconstruction loss $\mathcal{L}_{mel}$ from the Tacotron2 decoder like \cite{wang2018style}. Style encoder thus do not require any explicit style or prosody labels.
The modeling style in TTS is somewhat underdetermined, and training models with reconstruction loss alone are insufficient to disentangle content and style from other factors of variation.

The pathway of the style embedding constraints, which are represented by the blue dashed box in Fig. \ref{fig:model structure} (b), is trained using a loss between the predicted and ``target'' style embedding. 
An additional pre-trained encoder is used to extract the ``target'' embedding in the training process. Therefore, the first step is the style encoder pre-training, which can be simply treated as a neural GST training process. The pre-trained style encoder is trained using the same training sets. 
As in the case of TPCE-GST \cite{stanton2018predicting}, we stop the gradient flow to ensure that the style prediction error does not backpropagate through the pre-trained encoder.

The total loss with style embedding constraints is defined as
%\begin{center}
\begin{eqnarray}
\begin{aligned}
  \mathcal{L}_{total} = \mathcal{L}_{mel}+\mathcal{L}_{s},\\
   \mathcal{L}_{s} = \text{MSE}(\bm{E},\bm{E^{'}}) - \text{MI}(\bm{E},\bm{E^{'}}),
  %T_t &=& C_t * N_f/M_f      \\
  %T_l &=& C_l * F_l * N_f  \nonumber
  \end{aligned}
  \label{eq2}
\end{eqnarray}
%\end{center}
where $\mathcal{L}_{s}$ is the loss between the predicted style embedding $E$ and the target embedding $E^{'}$. Not only do predict and ``target'' tend to have the same value with low $\text{MSE}(\bm{E},\bm{E^{'}})$, but they also need to have strong mutual information $\text{MI}(\bm{E},\bm{E^{'}})$.

Mutual information (MI) measures the dependence of two random variables from the perspective of information \cite{hu2020unsupervised,belghazi2018mutual}. Given two random variables $X$ and $Y$, the MI $I(X; Y)$ between them is equivalent to the Kullback–Leibler (KL) divergence between their joint distribution, $P_{X,Y}$, and the product of marginals, $P_{X}P_{Y}$. 

The MI neural estimation (MINE) \cite{belghazi2018mutual} method constructs a lower bound of MI based on the Donsker-Varadhan representation of KL divergence via
%\begin{center}
\begin{eqnarray}
\begin{aligned}
  I(X;Y) \geq \hat{I}_{M}(X;Y) = \mathop{sup}\limits_{M}E_{{P}_{X,Y}}[M]-log(E_{P_{X}P_{Y}}[e^{M}]),
  \end{aligned}
  \label{eq4}
\end{eqnarray}
%\end{center}
where $M$ can be any function that forces the two expectations in the above equation to be finite. The authors in \cite{hu2020unsupervised} proposed the use of a deep neural network for $M$, which enables the MI between $X$ and $Y$ to be estimated by maximizing the lower bound in Eq. \ref{eq4} with respect to $M$ using gradient descent. 

The MI estimator function, $M$, is updated in each step of the training. The predicted style embeddings, $E$, and the ``target'' embeddings, $E'$, are used to train the MI estimator $M$. Then, the MI of $E$ and $E^{'}$, $\text{MI}(E,E^{'})$, can be estimated using the MI estimator $M$.
The MI estimator training process is summarized in Algorithm 1. 
%Because MI is always non-negative, we clip the estimated MI to zero if it is negative.
\begin{algorithm}[t]
	\renewcommand{\algorithmicrequire}{\textbf{Input:}}
	\renewcommand{\algorithmicensure}{\textbf{Output:}}
	\caption{Pseudocode for MI estimator training}
	\label{alg:1}
	\begin{algorithmic}[1]
		\REQUIRE Pairs of predicted and target embeddings $(E_{i},E^{'}_{i})$.
		\ENSURE $M$
		\STATE $M \gets $ initialization with random weights.
		\WHILE{$M$ not converged}
		\STATE Sample a mini-batch of $(E_{i},E^{'}_{i}),i=1,2,..b$.
		\STATE $E^{'}_{i} = $ random permutation of $E^{'}_{i}$.
		\STATE $\mathcal{L}_{mi} = \frac{1}{b}\sum_{i=1}^bM(E_{i},E^{'}_{i}) - log(\frac{1}{b}\sum_{i=1}^be^{M(E_{i},E^{'}_{i})})$
		\STATE $M=M+\epsilon {\Delta}_{M} \mathcal{L}_{mi}$
		\ENDWHILE
	\end{algorithmic}  
\end{algorithm}

\section{Experiments}
\subsection{Experimental Setup}
\label{sec:format}
\textbf{Datasets.} 
Experiments were performed on a high-quality dataset of English audiobook recordings featuring the voice of Catherine Byers (the speaker from the 2013 Blizzard Challenge). Some books contain very expressive character voices with high dynamic ranges, which are challenging to model. We selected five fairy tale books, and approximately 13,000 utterances were used for training  and validation. All speech data were downsampled to 22050 Hz. We trained an open-source WaveRNN \cite{pmlr-v80-kalchbrenner18a} vocoder using the same data to reconstruct waveforms from the Mel spectrogram.

\textbf{Method.} 
For the experiment, we built several TTS systems, as listed in Table \ref{tab:1}. 
We chose the vanilla Tacotron2 and TPSE-GST as our baseline systems (B1-B4) because GST-Tacotron requires either a reference signal or a manual selection of style token weights at the time of inference. Because the original TPSE-GST B2 only uses the mean square error (MSE) constraint, we also modified the TPSE model and added MI loss to it, which is B3-B4.

In addition, we included style embedding constrains C-MRTTS with and without MI loss to verify the effectiveness of the mutual information constraints.
The MRTTS was also built with and without attention to investigate whether or not the attention model was necessary.
The dimensionality of the style embeddings of all systems was 256.

%For our proposed MRETTS model, we have adopted two training methods: unsupervised U-MRETTS and supervised S-MRETTS.

\begin{table}[t]
  \caption{TTS systems used for our analysis. TPSE and Tacotron2 are baseline models, U-MRTTS and C-MRTTS represent the without or with style embedding constraint respectively. N means the number of multiple reference audios}
  \label{tab:1}
  \centering
  \begin{tabular}{|lccccl|}
\hline
         \multirow{2}*{\textbf{Method}} &\multirow{2}*{\textbf{Architecture}}  &
         \multirow{2}*{\textbf{Attention}} & \multicolumn{2}{c}{\textbf{Constraint}}
          & \multirow{2}*{\textbf{N}} \\ \cline{4-5}
         & &  &\textbf{MSE} &\textbf{MI} & \\
         \hline \hline
         B1 & Tacotron2 & N/A &N/A &N/A &N/A\\
         \hline \hline
         B2 & \multirow{3}{*}{TPSE-GST} &N/A &\checkmark &N/A &N/A \\
         B3  & &N/A &N/A  &\checkmark &N/A \\
         B4  & &N/A &\checkmark  &\checkmark &N/A \\\hline \hline
         P1 & \multirow{3}{*}{U-MRTTS} & N/A &N/A &N/A &3 \\
         P2 &  &\checkmark &N/A  &N/A &1 \\ 
         P3 &  &\checkmark &N/A  &N/A &3 \\ \hline \hline
         P4 &\multirow{7}{*}{C-MRTTS}  & N/A &\checkmark  &N/A &3 \\ 
         P5 &  & N/A &N/A   &\checkmark &3 \\
         P6 &  & N/A &\checkmark  &\checkmark &3 \\
         P7 &  &\checkmark  &\checkmark  &N/A &3 \\
         P8 &  &\checkmark  &N/A  &\checkmark &3 \\
         P9 &  &\checkmark  &\checkmark  &\checkmark &1 \\
         P10 &  &\checkmark  &\checkmark  &\checkmark &3 \\
\hline
\end{tabular}
\vspace{-0.5cm}
\end{table}

\textbf{Evaluation metrics.} 
In terms of subjective evaluation, the mean opinion score (MOS) was calculated on a scale from 1 to 5 with 0.5-point increments.
We also conducted an ABX test. The rating criterion was determined by answering the question ``Which one’s speaking style is closer to the target audio style?'' with one of three choices: (1) the first is better, (2) the second is better, and (3) neutral.
In all tests, 25 native listeners were asked to rate the performance of 50 randomly selected synthesized utterances from the test set. 

Because another main objective of the MRTTS algorithm is to reduce the content leakage of the generated speech and to improve the speech content quality, we objectively evaluated the performance by measuring the content quality using an ASR algorithm like \cite{hu2020unsupervised}. We adopted iFlytek’s online API \footnote{https://www.xfyun.cn/services/voicedictation} as the ASR system, and computed the word error rate (WER) as a metric for the content preservation ability of the model. 
We used a PyPi package called jiwer \cite{jiwer} to calculate the WER.

\subsection{Result and analysis}
\begin{table}[t]
  \caption{Mean opinion score (MOS) evaluations with 95\% confidence intervals computed from the t-distribution and WER for various systems.}
  \label{tab:2}
  \centering
  \setlength{\tabcolsep}{3mm}
  \begin{tabular}{|lccc|}
\hline
         \textbf{Method} & \textbf{Architecture} &\textbf{MOS} & \textbf{WER}  \\ \hline \hline
         B1 & Tacotron2 & $4.015\pm0.023$ & 28.2\% \\\hline \hline
         B2 & \multirow{3}{*}{TPSE-GST} &$4.175\pm0.016$ & 26.7\% \\
         B3  & & $4.177\pm0.022$ &26.3\%  \\
         B4 &  &$4.183\pm0.063$ &26.6\% \\\hline \hline
         P1 & \multirow{3}{*}{U-MRTTS}  &$4.011\pm0.074$ & 27.1\% \\
         P2 &  & $4.103\pm0.105$ & 21.2\% \\ 
         P3 &  & $4.292\pm0.035$ & 18.2\% \\ \hline \hline
         P4 &\multirow{7}{*}{C-MRTTS}  &$3.911\pm0.025$ & 30.2\%  \\ 
         P5 &  &$3.985\pm0.025$ & 29.3\% \\
         P6 &  &$3.997\pm0.031$ & 29.5\%  \\
         P7 &  &$4.123\pm0.015$  & 18.7\% \\
         P8 &  & $4.285\pm0.103$ & 18.4\%  \\
         P9 &  & $4.109\pm0.127$ & 19.1\%  \\
         \textbf{P10} &  &{$\bm{4.313\pm0.024}$}  & \textbf{17.9\%} \\ \hline \hline
         GT & N/A &$4.674\pm0.013$ &15.6\% \\\hline
\end{tabular}
\end{table}
\textbf{Objective evaluation of speech content quality.} 
We present our WER results in the last column of Table \ref{tab:2}, which shows that the proposed method produced a better WER than the baseline methods (a smaller WER indicates less content leakage). 
For the TPSE-GST model, the direct and only use of text to predict the style embedding easily caused the content information to be entangled into the style embedding.
Because we used multiple related reference audios as input instead of only target audio during training, style embedding is unlikely to contain more textual information compared to the baseline. Thus, our P3 and P10 methods can synthesize high-content quality speech and achieve a small WER in the ASR system.
The results also show that there is no significant difference between the WER produced by P3 and P10. Therefore, we infer that the improvement in speech content quality is mainly due to unpaired data rather than style embedding constraints.

\textbf{Subjective evaluation of speech naturalness.} 
The results of the MOS test are presented in the third column of Table\ref{tab:2}.
The proposed model P10 demonstrated a significantly better naturalness compared to the baseline models Tacotron2 and TPSE. 
This result shows the advantage of predicting style embeddings from multiple reference audios rather than from text.
Without the attention mechanism, the synthesized speech quality was significantly lower, especially for the C-MRTTS style embedding constraint models (P4-P6). 
P7, P8, and P10 yielded better results. However, for the baseline TPSE model, B3 and B4 are similar to B2, and there is no significant improvement.
On the one hand, MI is a better constraint for style embedding in our proposed model. On the other hand, MI does not bring significant gains to the TPSE model because MI constraints is limited by the style modeling ability of the TPSE model.
Furthermore, we also found that if only one audio was used as a reference, as in the cases of P2 and P9, the result was worse than when multiple audios were used, as in the cases of P3 and P10. 

%That means that our propose model could more easily fit the style embeddings obtained from multiple reference audios to the target style embeddings.
\begin{figure}[t]
  %\centering
  \includegraphics[width=0.48\textwidth]{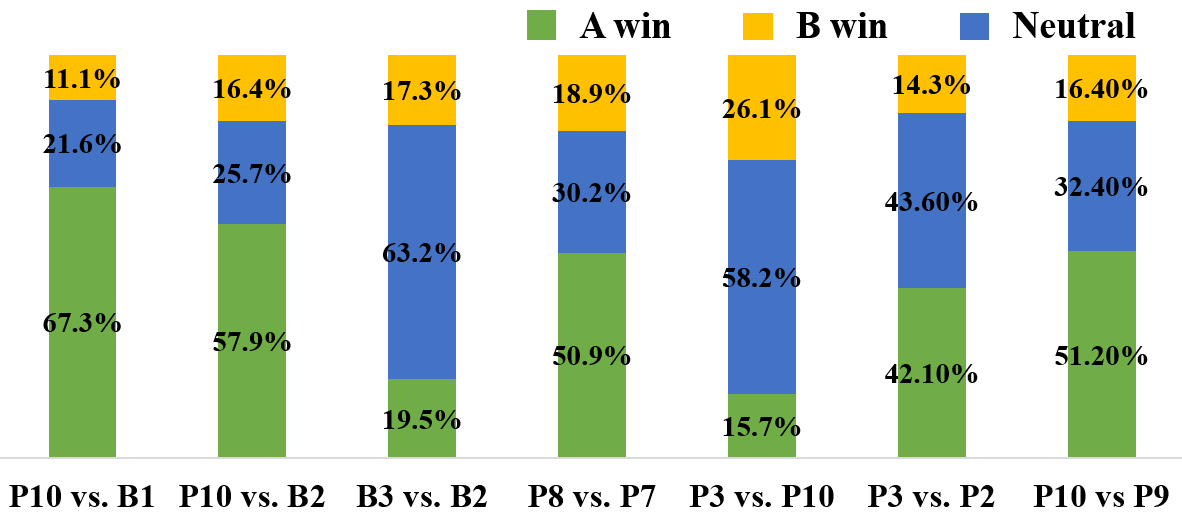}
  \vspace{-0.6cm}
  \caption{ABX test results for style similarity.}
  %\vspace{-0.4cm}
  \label{fig:abtest}
  \setlength{\belowcaptionskip}{0.cm}
  \vspace{-0.6cm}
\end{figure}

\textbf{Subjective evaluation of ABX test for style similarity.} Fig. \ref{fig:abtest} shows the results of ABX test.
As expected, a gap between our proposed model P10 and the baseline models B1 and B2 is visible. This shows that the proposed MRTTS model can produce better latent style representations, which results in better style similarity. 
The results also show that raters had a strong preference for P3 and P10.
That means that the style constraint methods we proposed can perform well in style models.
The results of the ABX test for the TPSE baseline model show that B3 was similar to B2, and there is no significant difference.
However, comparing P7 and P8, the style embedding constraint with mutual information improved the style model performance.
This proves once again that the use of MI cannot compensate for the shortcomings of the TPSE model.
In addition, we also found that if only one audio was used
as the reference, as in the cases of P2 and P9, the style similarity was worse than when 
multiple audios were used, as in the cases of P3 and P10. 
In other words, multiple audios contain richer style information than a single audio.
\begin{comment}
\textbf{Analysis of the style embeddings space.} We use t-SNE to visualize the style embeddings learned from various systems like Fig. \ref{fig:tsne}.
Note that ``target embedding" were extracted from pre-train style encoder using target audios.
From the visualization it is easy to see style embeddings from P3 and P10 are much close to target embeddings than B2.
Hence, the method of using multiple reference audios in proposed MRTTS is better than the method of using text in TPSE to model the speech style better.
In addition, P3 have a lager intra-cluster distance than P10.
This shows that without style constraint may make learning styles more scattered and diverse.
What's more, compared with P10, the embedding of P9 model is far from target embedding than that of P9. This result shows that it is easier to fit the target embeddings with multiple related reference audio than just one reference.
\end{comment}

\begin{figure}[t]
  \centering
  \includegraphics[width=0.46\textwidth]{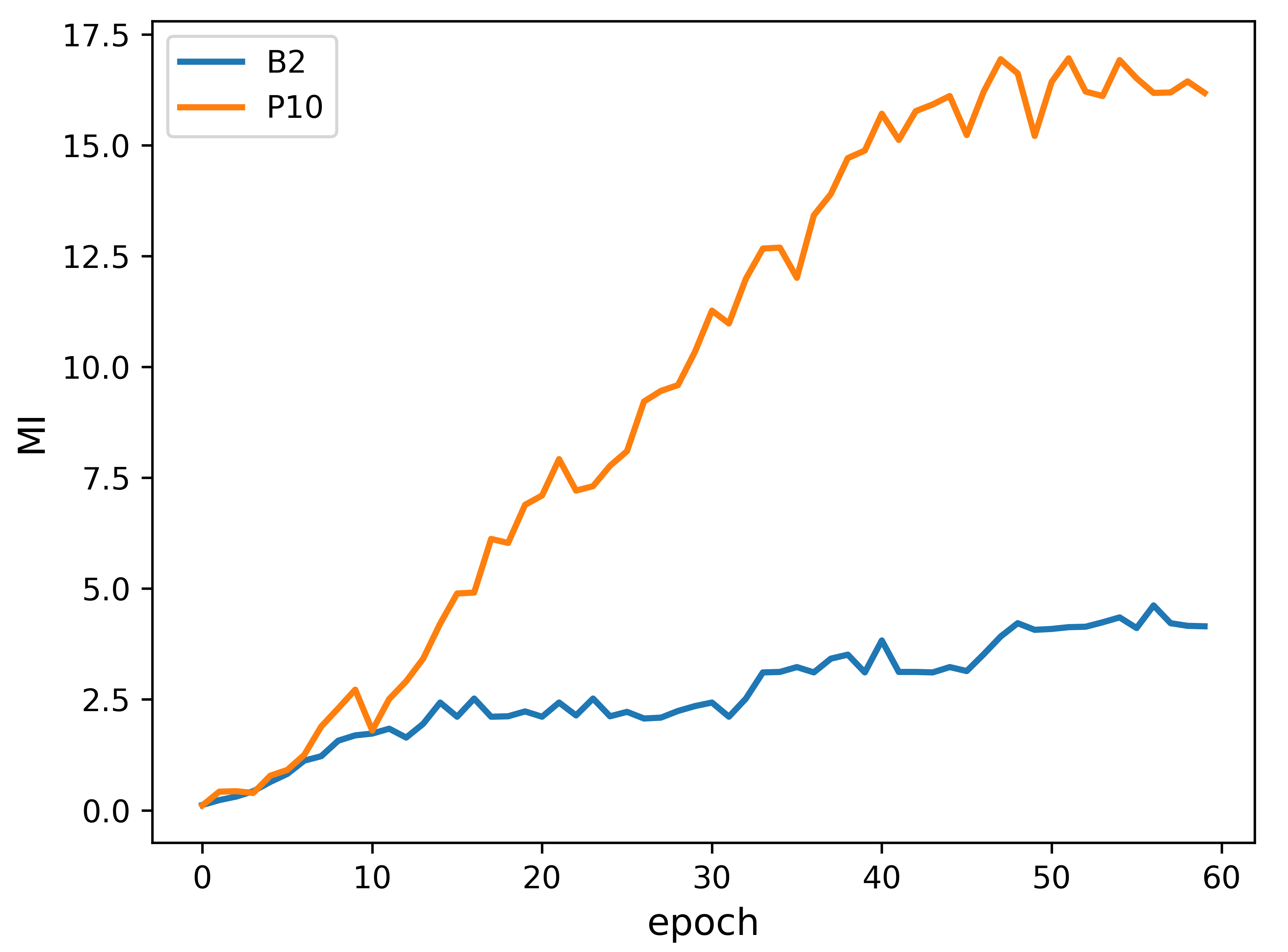}
  \vspace{-0.3cm}
  \caption{Estimated MI value in the training process.}
  \label{fig:tsne}
  \setlength{\belowcaptionskip}{0.cm}
  \vspace{-0.5cm}
\end{figure}

\textbf{Mutual Information Evaluation.} 
We estimate the mutual information between the two random variables (i.e. the predict style embedding and the ``target'' embedding) from our trained model (with frozen weights) using the MINE algorithm, and the result is illustrated in Fig. \ref{fig:tsne}. As expected, P10 with style embedding constraint have much high MI value than the B2. We assume this is because the style embedding constraints in MI model may improve the mutual information between predict and ``target'' style embedding and consequently increase the MI value.

In summary, using multiple reference audios (rather than only the target audio) as input reference can more accurately model latent style representations. In addition, the style constraints of MI can improve the speech style similarity. Both objective and subjective evaluations showed that our proposed MRTTS can synthesize more intelligible and natural speech. 
We present synthetic samples at \url{https://gongchenghhu.github.io/ICASSP2022_demo/}.

%tsne图，三个embedding图，tpse图，MRETTS图
\section{Conclusion and future work}
\label{sec:prior}
In this paper, we proposed a MRTTS model that uses multiple reference audios and style embedding constraints for the synthesis of expressive speech. 
The multiple reference audios were automatically selected using the sentence similarity determined by BERT. 
In addition, we considered the MI between style embeddings as constraints.
The experimental results showed that the proposed model can improve speech naturalness and content quality, and that it can outperform the baseline model according to ABX preference tests of style similarity.  In the future, we  aim  to build a fine-grained model that learns variable-length style information from multi-reference audios.

\vfill\pagebreak

% References should be produced using the bibtex program from suitable
% BiBTeX files (here: strings, refs, manuals). The IEEEbib.bst bibliography
% style file from IEEE produces unsorted bibliography list.
% -------------------------------------------------------------------------
\bibliographystyle{IEEEbib}
\bibliography{strings,refs}

\end{document}